\documentclass[pra,showpacs,twocolumn]{revtex4}
\usepackage{amsfonts}
\usepackage{amsmath}
\usepackage{amssymb}
\usepackage{graphicx}

\begin{document}

\title{Relationship between spin squeezing and single-particle coherence
in two-component Bose-Einstein condensates with Josephson coupling}
\author{G. R. Jin$^{1,2}$ and C. K. Law$^{1}$}
\affiliation{$^1$Department of Physics and Institute of Theoretical
Physics, The Chinese University of Hong Kong, Shatin, NT,
Hong Kong SAR, China\\
$^2$ Department of Physics, Beijing Jiaotong University, Beijing
100044, China}
\date{\today }

\begin{abstract}
We investigate spin squeezing of a two-mode boson system with a
Josephson coupling. An exact relation between the squeezing and the
single-particle coherence at the maximal-squeezing time is
discovered, which provides a more direct way to measure the
squeezing by readout the coherence in atomic interference
experiments. We prove explicitly that the strongest squeezing is
along the $J_z$ axis, indicating the appearance of atom
number-squeezed state. Power laws of the strongest squeezing and the
optimal coupling with particle number $N$ are obtained based upon a
wide range of numerical simulations.
\end{abstract}

\maketitle

\section{Introduction}

Spin squeezing is a nonclassical effect of collective spin systems
\cite{Kitagawa,Wineland,Kuzmich,You}, showing reduced spin
fluctuation in one certain spin component normal to the mean spin.
Kitagawa and Ueda proposed spin squeezing generated by the
self-interaction Hamiltonian $H_{1}=2\kappa J_{z}^{2}$, due to the
so-called one-axis twisting (OAT) effect \cite{Kitagawa}. The
OAT-type spin squeezing could be realized in weakly interacting
Bose-Einstein Condensate (BEC) \cite{Sorensen}, or atomic ensemble
in a dispersive regime \cite{Takeuchi}. The self-interaction $H_{1}$
also leads to phase diffusion of the BEC \cite{Lewenstein}, which
indicates a decay of single-particle coherence
\cite{Orzel,Schumm,Chuu,Jo,Artur,Vardi}.

Beyond the OAT model, an Josephson-like coupling (JLC) term $\Omega
J_{x}$ was added to the Hamiltonian $H_{1}$ with purpose to
coherently control the phase diffusion \cite{Vardi} and the spin
squeezing \cite{Bigelow,Law,Jin07}. It was shown that the JLC model
[see Eq.(\ref{H2})] results in strong reduction of spin fluctuation
along the $z$ (i.e., $J_z$) axis, provided that the additional field
is tunned optimally \cite{Law}. We found the maximal-squeezing time
$t_0$ of the JLC model, and proposed a simple scheme to store the
strongest squeezing along the $z$ axis for a long time \cite{Jin07}.
So far, there remain certain questions unsolved: Is there any
relation between the squeezing and the single-particle coherence? In
addition, to what degree can the strongest squeezing reach in the
JLC model? The first question is important because it relates to
measurement of the squeezing.

In this paper, we present an exact relation between the squeezing
and the coherence by solving the Heisenberg equation. Our results
show that local minima of the squeezing and the coherence occur
simultaneously for the coupling $\Omega$ larger than its optimal
value $\Omega_0$. Unlike the OAT scheme, where number variance
$\Delta J_{z}$ is time-independent, we prove explicitly that the
squeezing at time $t_{0}$ is along the $z$ axis in the JLC model
\cite{Note}. The strongest squeezing obeys the power law
$\xi_{0}=\Delta J_{z}(t_0)/\sqrt{j/2}\propto N^{-1/3}$, which can be
measured by readout the single-particle coherence through the
visibility of the interference fringe
\cite{Orzel,Schumm,Chuu,Jo,Artur}.

Our paper is organized as follows. In Sec. II, we introduce
theoretical model and derive some formulas for the single-particle
coherence and the squeezing parameter. In Sec. III, quantum dynamics
of the coherence and the squeezing are investigated for the OAT and
the JLC models, respectively. In Sec. IV, we present exact relation
between the coherence and the spin squeezing at the
maximal-squeezing time $t_0$. In Sec. V, power rules of the optimal
coupling and the strongest squeezing as a function of particle
number $N$ are investigated based upon a wide range of numerical
simulations. Moreover, we compare numerical result of $t_0$ for the
optimal coupling case with its analytic solution. Finally, a summary
of our paper is presented.

\section{Theoretical model and some formulas}

To begin with, we consider a two-component weakly-interacting BEC
consisting of $2j$ atoms in two hyperfine states $|1\rangle$ and
$|2\rangle$ coupled by a radio-frequency (or microwave) field
\cite{Hall,Stenger}. A tightly-confined BEC can be described by the
JLC Hamiltonian ($\hbar=1$) \cite{TMA}
\begin{equation}
H_{2}=\Omega J_{x}+2\kappa J_{z}^{2},  \label{H2}
\end{equation}
where the angular momentum operators $J_{+}=(J_{-})^{\dag
}=\hat{a}_{2}^{\dag }\hat{a}_{1}$, $J_{z}=(\hat{a}_{2}^{\dag
}\hat{a}_{2}-\hat{a}_{1}^{\dag }\hat{a}_{1})/2$ obey the SU(2) Lie
algebra. The total particle number $N=\hat{a}_{1}^{\dag
}\hat{a}_{1}+\hat{a}_{2}^{\dag }\hat{a}_{2}$ is a conserved
quantity. In Eq.(\ref{H2}), we have neglected the term proportional
to $J_z$ by assuming equal intraspecies atom-atom interaction
strengthes \cite{Sorensen}. The Rabi frequency $\Omega$ can be
controlled by the strength of the external field. The
self-interaction term $2\kappa J_{z}^{2}$ leads to spin squeezing,
which is quantified by a parameter \cite{Kitagawa}:
\begin{equation}
\xi =\frac{\sqrt{2}(\Delta J_{\mathbf{n}})_{\min }}{j^{1/2}},  \label{xi}
\end{equation}%
where $j=N/2$ and $(\Delta J_{\mathbf{n}})_{\min }$ represents the
minimal variance of a spin component $J_{\mathbf{n}}=\vec{J}\cdot
\mathbf{n}$ normal to the mean spin $\langle \vec{J}\rangle$. The
coherent spin state (CSS), defined formally as \cite{CSS}
\begin{equation}
\left\vert \theta ,\phi \right\rangle =e^{-i\theta (J_{x}\sin \phi
-J_{y}\cos \phi )}|j,-j\rangle  \label{CSS}
\end{equation}%
has the minimal variance $(\Delta J_{\mathbf{n}})_{\min
}=\sqrt{j/2}$ and $\xi =1$. Therefore, a state is called spin
squeezed state if its variance is smaller than that of the CSS, i.e.
$\xi<1$. Besides the squeezing, the self-interaction $2\kappa
J_{z}^{2}$ also leads to the phase diffusion, which indicates a
decay of the single-particle coherence. Such a kind of coherence is
measured by off-diagonal elements of the single-particle density
matrix $\rho_{ij}^{(1)}=\langle a_{i}^{\dag }a_{j}\rangle/N$ with
$i,j=1,2$. Formally, one introduces the first-order temporal
correlation function \cite{Vardi}:
\begin{equation}
g_{12}^{(1)}=\frac{\left\vert \rho _{12}^{(1)}\right\vert }{\sqrt{\rho
_{11}^{(1)}\rho _{22}^{(1)}}}\equiv \frac{\left\vert \left\langle
J_{+}\right\rangle \right\vert }{\sqrt{j^{2}-\left\langle J_{z}\right\rangle
^{2}}},  \label{coherence}
\end{equation}%
which is observable in experiments by extracting the visibility of
the Ramsey fringes \cite{Orzel,Schumm,Chuu,Jo,Artur}. One of the
goals of this paper is thereby to present the relation between the
squeezing $\xi$ and the first-order coherence $g_{12}^{(1)}$.

Let us first examine the exact numerical solutions of the
time-dependent Schr\"{o}dinger equation governed by the JLC
Hamiltonian $H_{2}$. We consider that the spin system starts from
the lowest eigenvector of $J_{x}$, $|j,-j\rangle_{x}=e^{-i\pi
J_{y}/2}|j,-j\rangle$, a particular CSS, Eq.~(\ref{CSS}), with
$\theta =\pi /2$ and $\phi=\pi$. Such an experimentally realizable
state can be prepared by applying a two-photon $\pi/2$ pulse to the
ground state $|j,-j\rangle$ with all the atoms in the internal state
$|1\rangle$ \cite{Sorensen,Artur,Hall}. The spin state at arbitrary
time $t$ can be expanded as: $|\Psi \rangle =\sum_{m}c_{m}\left\vert
j,m\right\rangle$, and the amplitudes $c_{m}$ obey
$i\dot{c}_{m}=\varepsilon _{m}c_{m}+X_{-m}c_{m+1}+X_{m}c_{m-1}$,
where $\varepsilon _{m}=2\kappa m^{2}$, $X_{m}=\frac{\Omega
}{2}\sqrt{(j+m)(j-m+1)}$ with $X_{-j}=0$. The amplitudes of the
initial CSS are $c_{m}(0)=\langle j,m|j,-j\rangle
_{x}=\frac{(-1)^{j+m}}{2^{j}}\binom{2j}{j+m}^{1/2}$. Obviously,
$c_{-m}(0)=c_{m}(0)$ for even $N$ and $c_{-m}(0)=-c_{m}(0)$ for odd
$N$, which gives the expectation value $\langle J_{z}(0)\rangle =0$
and the variance $\langle J_{z}^{2}(0)\rangle =j/2$. Note that some
references adopt the Hamiltonian $H_3=-\Omega J_x+2\kappa J^2_z$ to
investigate the BEC in a double-well potential \cite{TMA}, which
corresponds to $H_2$ for $\Omega<0$ case. In this work, we consider
only positive $\Omega$ and $\kappa$ by assuming repulsive atom-atom
interactions. If either $\Omega$ or $\kappa$ is negative, our
results remain valid by using initial state $|j, j\rangle_x$
\cite{Hines}.

Since $c_{-m}(0)=\pm c_{-m}(0)$ and $X_{\pm m}=X_{\mp m+1}$, we
introduce linear combinations of the amplitudes $p_{m}^{(\pm
)}=c_{m}\pm c_{-m}$. For even $N$ case, one can derive a closed set
of equations for $p_{m}^{(-)}$. However, all $p_{m}^{(-)}(0)=0$ lead
to $p_{m}^{(-)}(t)=0$, thus $c_{-m}(t)=c_{m}(t)$. As a result,
dynamical evolution of the even $N$ system is determined solely by
the equations of the amplitudes $p_{m}^{(+)}$ with $m=0,1,...,j$.
Similarly, for the odd $N$ case, we obtain $
p_{m}^{(+)}(0)=p_{m}^{(+)}(t)=0$, i.e., $c_{-m}(t)=-c_{m}(t)$.
Quantum dynamics of the odd $N$ system depends on the equations of
$p_{m}^{(-)}$. The above processes have certain advantages to: (i)
reduce the total Hilbert space dimension from $2j+1$ to $j+1$ (even
$N$) or $j+1/2$ (odd $N$); (ii) Since $c_{-m}=\pm c_{m}$, we obtain
$\langle J_{y}\rangle =\langle J_{z}\rangle =0$ and $\langle
J_{x}\rangle =\langle J_{+}\rangle \neq 0$, i.e., the mean spin is
always along the $x$ axis. Actually $\langle J_{+}\rangle $ is a
real function; (iii) The correlation function is simplified as
$g_{12}^{(1)}=j^{-1}\left\vert \left\langle J_{x}\right\rangle
\right\vert $, and the spin component normal to the mean spin is
$J_{\mathbf{ n}}=J\cdot \mathbf{n}=J_{y}\sin \theta -J_{z}\cos
\theta $. The variance of $ J_{\mathbf{n}}$ is $(\Delta
J_{\mathbf{n}})^{2}=\langle J_{\mathbf{n}}^{2}\rangle -\langle
J_{\mathbf{n}}\rangle ^{2}=(C-A\cos 2\theta -B\sin 2\theta )/2$,
where $A=\langle J_{y}^{2}-J_{z}^{2}\rangle $, $B=\langle
J_{z}J_{y}+J_{y}J_{z}\rangle $, and $C=\langle
J_{y}^{2}+J_{z}^{2}\rangle $. From the relation $\left. \partial
_{\theta }(\Delta J_{\mathbf{n}})^{2}\right\vert _{\theta _{\min
}}\equiv 0$, we get $\tan (2\theta _{\min })=B/A$ and the minimal
variance $(\Delta J_{\mathbf{n}})_{\min }^{2}=\left(
C-\sqrt{A^{2}+B^{2}}\right)/2$.

\section{quantum dynamics of the coherence and the squeezing}

\begin{figure}[htbp]
\begin{center}
\includegraphics[width=6cm, angle=270]{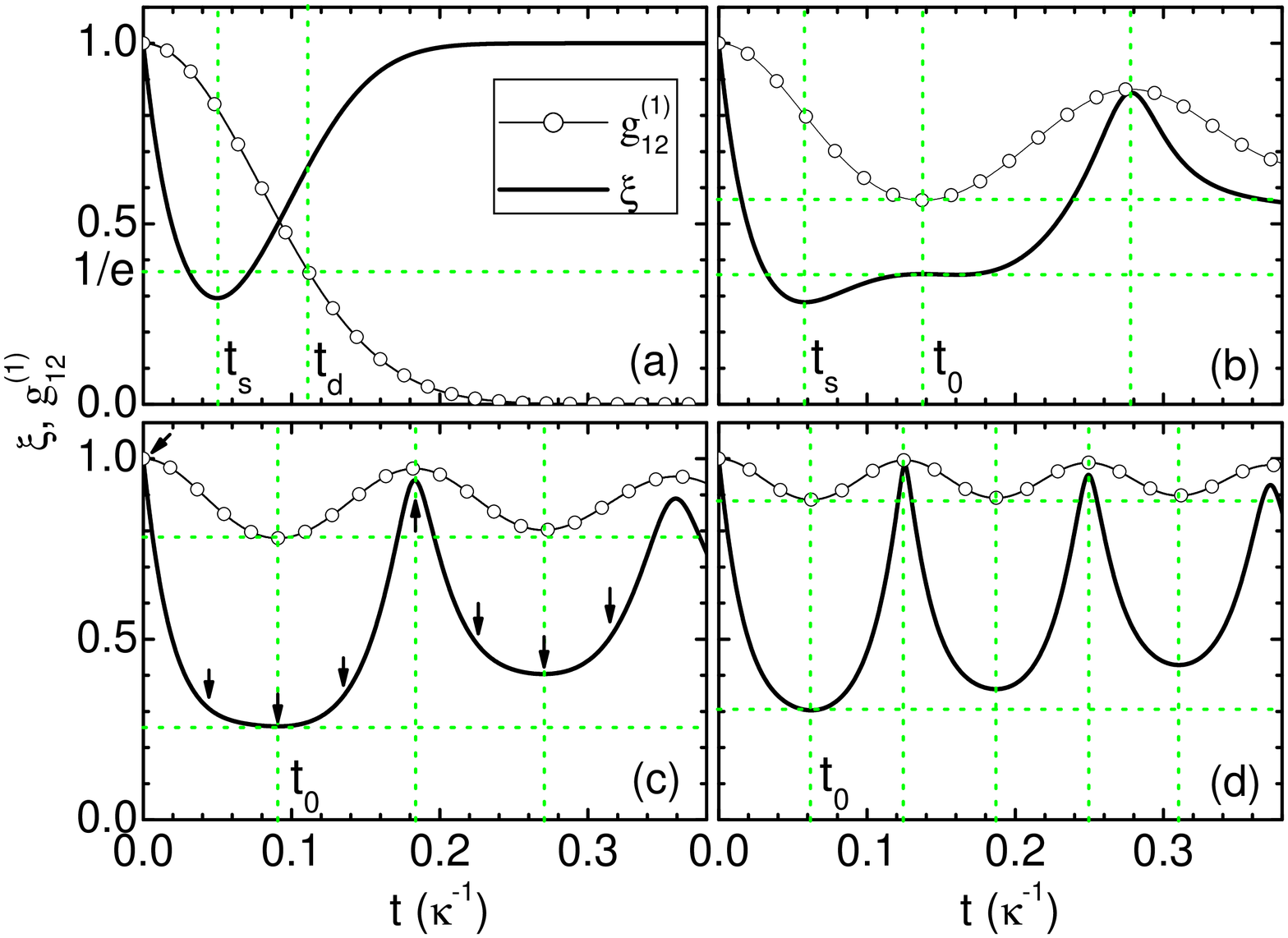} \vskip -0.2cm
\caption{Time evolution of the squeezing parameter $\xi$ (solid
thick lines) and the single-particle coherence $g_{12}^{(1)}(t)$
(thin lines with open circles) for $N=40$ and various Rabi
frequencies: (a) $\Omega=0$, (b) $\Omega=2$, (c)
$\Omega=\Omega_0=4.2405$ (optimal coupling), (d) $\Omega=8$. Time
$t$ is in units of $\kappa^{-1}$, and the units of $\Omega$ is
$\kappa$.}\label{fig1}
\end{center}
\end{figure}

The OAT model $H_{1}$ (i.e., $\Omega =0$) can be solved exactly in
Heisenberg picture \cite{Kitagawa}, with its analytic results:
$A=(j/2)(j-1/2)\left[1-\cos ^{2j-2}(4\kappa t)\right]$,
$B=-j(2j-1)\sin (2\kappa t)\cos^{2j-2}(2\kappa t)$, and $C=j+A$ due
to the time-independent variance $\langle J_{z}^{2}\rangle=j/2$. The
strongest (optimal) squeezing $\xi_{0}=\xi(t_{s})\simeq
(4/3)^{1/6}N^{-1/3}$ occurs at time $t_{s}\simeq6^{1/6}N^{-2/3}/2$.
The single-particle coherence $g_{12}^{(1)}(t)=\cos^{N-1}(2\kappa
t)\simeq e^{-(t/t_{d})^{2}}$ with the phase-diffusion time $\kappa
t_{d}=(2N)^{-1/2}$. Obviously,
$g_{12}^{(1)}(t_{d})=e^{-1}g_{12}^{(1)}(0)=1/e$. The coherence
$g_{12}^{(1)}(t)$ has been measured in experiment by extracting the
visibility of the Ramsey fringe \cite{Artur}. As shown in
Fig.~\ref{fig1}(a), the optimal squeezing occurs within the
coherence time due to $t_{s}<t_{d}$. Moreover, the coherence
$g_{12}^{(1)}$ decays to zero at $t_{0}=\pi/(4\kappa)$, and recover
to unity at $2t_{0}$ [not shown in Fig.~\ref{fig1}(a), see
Refs.\cite{Lewenstein,Vardi}].

Except for $N=2,3$ cases, the JLC model ($\Omega \neq 0$) can
\textit{not} be solved exactly \cite{Law,Jin07,Agarwal}. Numerical
simulations of the single-particle coherence $g_{12}^{(1)}(t)$ and
the squeezing $\xi(t)$ are presented in Fig.~\ref{fig1}(b)-(d) for
$N=40$ and various $\Omega $. Similar with the OAT case, the
coherence $g_{12}^{(1)}(t)$ collapses to its local minimum at
$t_{0}$ then revives partially at about $2t_{0}$, and the maximal
squeezing occurs at $t_{s}$ for a small coupling $\Omega =2\kappa $
[Fig.~\ref{fig1}(b)]. Two time scales $t_{0}$ and $t_{s}$ tend to
merge with the increase of $\Omega$. For $\Omega \geq \Omega _{0}$,
both the coherence $g_{12}^{(1)}$ and the squeezing $\xi$ reach
local minima at the same time $t_{0}$ [Fig.~\ref{fig1}(c) and (d)].
Here $\Omega_{0}$ is the optimal coupling to produce the strongest
squeezing in the JLC model, such as $\Omega_{0}=4.2405\kappa$ for
$N=40$.

It should be mentioned that the spin state at $t_0$ exhibits a very
sharp probability distribution and a strong reduction of the number
variance $\Delta J_z$ \cite{Jin07}. The loss of the coherence (or
visibility) as an evidence of the number squeezing at time $t_0$, as
shown in Fig.~\ref{fig1}(b)-(d), has been observed by Orzel et al.
\cite{Orzel}. Moreover, our results show that the phase diffusion is
suppressed due to the appearance of number-squeezed state at $t_0$,
consistent with the experimental observations \cite{Jo}. In
experiments so far, however, the number squeezing were detected
through the observation of increased phase fluctuation $\Delta
\phi$, or through an increased phase-diffusion time \cite{Gerbier}.
In fact, the number fluctuation has a nontrivial relation with the
phase fluctuation, complex even for a single-mode light field
\cite{Pegg}. Consequently, a more direct way to measure the number
fluctuation $\Delta J_z$ is necessary.

\section{Exact relation between the coherence and the squeezing at the maximal-squeezing time}

There exists an exact relation between the coherence $g_{12}^{(1)}$
and the squeezing $\xi$ at time $t_{0}$. To see this, let us examine
the Heisenberg equations: $\dot{J}_{x}=-2\kappa \left(
{J_{z}J_{y}+J_{y}J_{z}}\right)$ and $\dot{J} _{z}=\Omega J_{y}$. The
first equation gives the relation between the coherence
$g_{12}^{(1)}$ and the squeezing angle $\theta_{\min}$
\begin{equation}
\frac{d}{dt}g_{12}^{(1)}=2\kappa Aj^{-1}\tan (2\theta _{\min
}).\label{gq}
\end{equation}%
As $g_{12}^{(1)}$ reaches its local minimum at $t_{0}$, $
(dg_{12}^{(1)}/dt)_{t_{0}}\equiv 0$, then the squeezing angle
$\theta _{\min}=0$ (or $\pi$) provided $A\neq 0$. Combining the two
Heisenberg equations, we obtain further $dJ_{z}^{2}/dt=\Omega
({J_{z}J_{y}+J_{y}J_{z}})=-(\Omega /2\kappa )dJ_{x}/dt$, which
yields $\langle J_{z}^{2}\rangle =\lambda -\Omega \langle
J_{x}\rangle/(2\kappa)$ with the integral constant $\lambda$. For
the initial CSS $|j,-j\rangle_{x}$, we have $\lambda =j(1-\Omega
/\kappa)/2$. At time $t_{0}$, $B=0$ and $A>0$, and the minimal
variance $(\Delta J_{\mathbf{n}})_{\min }^{2}=(C-A)/2\equiv \langle
J_{z}^{2}\rangle$, i.e., the squeezing along the $z$ axis
\cite{Note}. As a result, we obtain a simple relation between
$g_{12}^{(1)}$ and $\xi$
\begin{equation}
\xi ^{2}(t_{0})=\frac{2\left\langle J_{z}^{2}(t_{0})\right\rangle
}{j}=1-\frac{\Omega }{\kappa }\left[ 1-g_{12}^{(1)}(t_{0})\right] ,
\label{gxi}
\end{equation}%
which is valid for arbitrary $\Omega $. For instance $\Omega=0$,
$g_{12}^{(1)}=0$ and $\xi=1$ at $t_{0}=\pi/(4\kappa)$; while for
$\Omega
>N\kappa$, from Eq.~(\ref{gxi}) we get $g_{12}^{(1)}(t_{0})\simeq \xi
(t_{0})\simeq 1$; two trivial results due to weak squeezing.
Hereafter, we focus on the coupling around its optimal value $\Omega
_{0}$, with which the strongest squeezing $\xi_{0}=\xi
(t_{0},\Omega_{0})$ can be obtained at time $t_{0}$. According to
Eq.~(\ref{gxi}), one can measure $\xi_{0}$ by readout the coherence
$g_{12}^{(1)}$ in atomic interference experiments
\cite{Orzel,Schumm,Chuu,Jo,Artur}.

\begin{figure}[htbp]
\begin{center}
\includegraphics[width=6cm, angle=270]{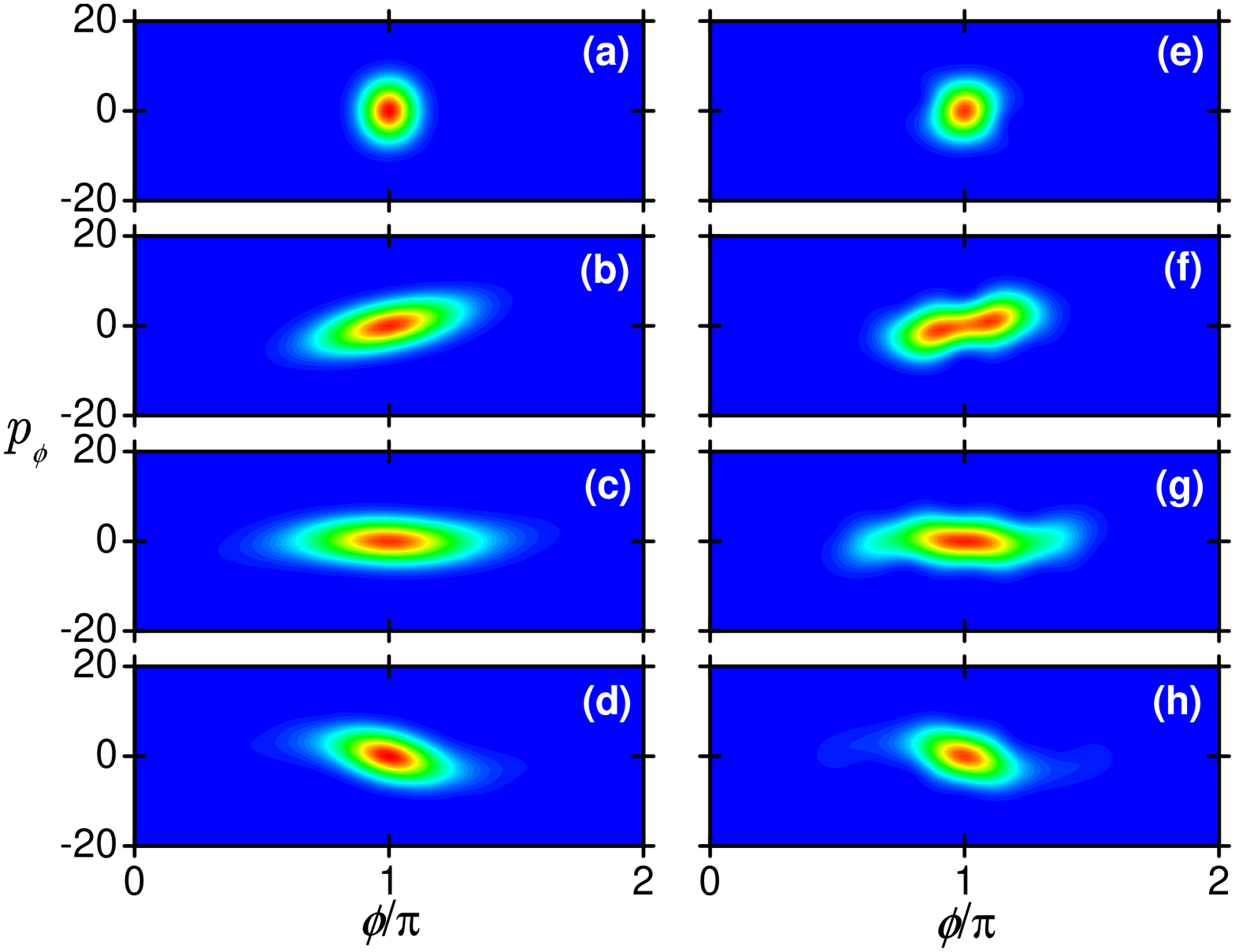} \vskip -0.2cm
\caption{(color online) Husimi $Q$ function in the phase space
$(\phi, p_{\phi})$ for various times: (a) $t=0$, (b) $t=0.045$, (c)
$t=t_0=0.09064$ (the optimal squeezing), (d) $t=0.135$, (e)
$t=0.183127$, (f) $t=0.225 $, (g) $t=0.26988$, and (h) $t=0.315$,
corresponding to the times indicated by the arrows in
Fig.~\ref{fig1}(c), respectively. Other parameters are $N=40$,
$\Omega=\Omega_0=4.2405\kappa$, and time $t$ is in units of
$\kappa^{-1}$. } \label{fig2}
\end{center}
\end{figure}

The time scale $t_{0}$ can be obtained based upon the phase model
\cite{Jin07}. By replacing $J_{z}\rightarrow p_{\phi }=-i\partial
_{\phi }$ and $J_{x}\rightarrow j\cos \phi $, the JLC Hamiltonian
$H_{2}$ can be rewritten as $H_{\phi }=2\kappa p_{\phi }^{2}+\Omega
j\cos \phi $ \cite{phase model}, where $\phi$ is relative phase
between two bosonic modes. The phase model allows us to treat the
spin system as a simple classical pendulum oscillating around the
minimum of the Mathieu potential $\cos \phi$, i.e., $\phi =\pi$. In
the Josephson regime $1<\Omega /\kappa \ll N$, the pendulum rotates
in the phase space ($\phi ,p_{\phi}$) with the effective frequency
$\omega _{\text{eff}}=\sqrt{2\kappa \Omega N}$ \cite{Jin07,Leggett}.
To illustrate this motion, we calculate the Husimi $Q$ function
\begin{equation}
Q(\theta ,\phi )=\left\vert \langle \theta ,\phi |\Psi (t)\rangle
\right\vert ^{2}
\end{equation}%
in the phase space ($\phi ,p_{\phi}$), where $p_{\phi }=-j\cos
(\theta)$ describes the population imbalance of the two modes, and
the polar angle $\phi$ represents the relative phase
\cite{Trimborn}. The CSS $|\theta ,\phi \rangle$ is given in
Eq.~(\ref{CSS}). As shown in Fig.~\ref{fig2}(a), the $Q$ function is
a circle for the initial CSS, which represents Poisson distribution
of the number variance $(\Delta J_{z})^{2}$ and the phase
uncertainty $(\Delta \phi)^{2}$ \cite{Artur}. As time increases, it
becomes an elliptic shape [Fig.~\ref{fig2}(b)], rotating clockwise
in the phase space. After a duration $t_{0}$, the ellipse elongates
horizontally corresponding to the optimal squeezing of $(\Delta
J_{z})^{2}$ [Fig.~\ref{fig2}(c)]. It seems reasonable to suppose
that the motion of the ellipse is consistent with that of the
pendulum. In fact, the trajectory of the pendulum is just passing
through the horizontal axis ($p_{\phi }=0$) at time $T/4$, where
$T=2\pi /\omega _{\text{eff}}$ is the period of the pendulum. As a
result, we get the maximal-squeezing time \cite{Jin07}
\begin{equation}
\kappa t_{0}\simeq \kappa \frac{T}{4}=\frac{\pi
}{2}\sqrt{\frac{\kappa }{2\Omega N}},  \label{t0_JLC}
\end{equation}%
which is valid for large $N$ ($\geq 10^{3}$). As shown in
Fig.~\ref{fig2}(e)-(h), for $t\geq 2t_{0}$ ($\simeq T/2$) the $Q$
functions almost recover to original shapes, due to partial revival
of the squeezing $\xi$ and the coherence $g_{12}^{(1)}$ [see
Fig.~\ref{fig1}(c)].

\section{Optimal coupling and the strongest squeezing}

\begin{figure}[tbph]
\begin{center}
\includegraphics[width=8.5cm, angle=0]{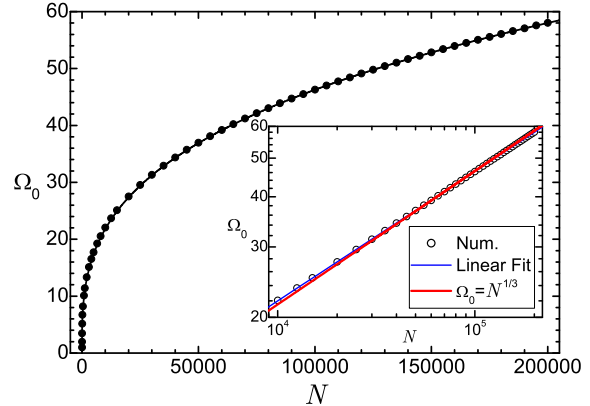} \vskip -0.6cm
\caption{(color online) Numerical simulation of the optimal coupling
$\Omega_{0}$ as a function of $N$ (solid circles) in the normal
scale. The solid line is the fit to power law $\Omega_{0}=aN^{b}$
with $a=1.07827$ and $b=0.32655$. The inset: comparison linear fit
(blue line) of numerical results (open circles) with $
\Omega_{0}=N^{1/3}$ (red line) in the log-log scale. $\Omega _{0}$
is in units of $\kappa$.} \label{fig3}
\end{center}
\end{figure}

To create the strongest reduction of $(\Delta J_{z})^{2}$ as shown
in Fig.~\ref{fig2}(c), we need to determine the optimal coupling
$\Omega_{0}$ as a function of particle number $N$. Note that the
optimal squeezing occurs at $t_s$ for the OAT and $t_0$ for the JLC,
respectively. For large $N$, the latter time scale should be
comparable with the former one as $\Omega=\Omega_{0}$. Such a
non-rigorous comparison enables us to suppose a power law as
$\Omega_{0}/\kappa \simeq N^{1/3}$. Numerical solution of
$\Omega_{0}$ is presented in Fig.~\ref{fig3} for $N$ up to $2\times
10^{5}$. We fit the data as $\Omega_{0}/\kappa=aN^{b}$ and find the
power-exponent $b=0.32655$, very close to the expected value $1/3$.
From the inset of Fig.~\ref{fig3}, we also find that the larger
number $N$ is adopted, the better fit is obtained.

\begin{figure}[tbph]
\begin{center}
\includegraphics[width=8.5cm, angle=0]{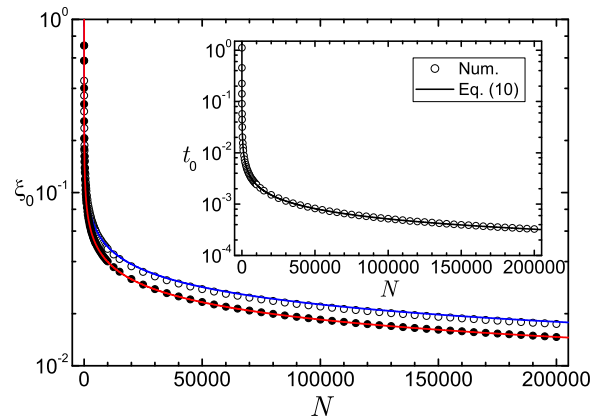} \vskip -0.6cm
\caption{(color online) The optimal squeezing $\xi_{0}$ for the OAT
(open circles) and the JLC (solid circles) as a function of $N$. The
blue line is given by $\xi_{0}\simeq (4/3)^{1/6}N^{-1/3}\simeq
1.0491N^{-1/3}$; the red line is a fitting curve as $\protect\xi_0=
0.8578N^{-1/3}$. The inset: numerical simulations of $t_{0}$ (open
circles), and the analytic result of Eq.~(\ref{t0_JLC}) with $\Omega
_{0}/\kappa=N^{1/3}$ (solid line) as a function of $N$ for optimal
squeezing. The time $t_0$ is in units of $\kappa^{-1}$.}
\label{fig4}
\end{center}
\end{figure}

In Fig.~\ref{fig4}, we investigate the optimal squeezing $\xi_{0}$
as a function of $N$. The fitting result is $\xi_{0}\simeq
0.8578N^{-1/3}$, which is slightly smaller than the OAT result $\xi
_{0}\simeq (4/3)^{1/6}N^{-1/3}\simeq 1.0491N^{-1/3}$. Small
difference of $\xi_{0}$ between the OAT and the JLC does not
deteriorate the advantages of the latter scheme. In fact, there is
no number squeezing in the OAT model due to $[J_z^2, H_1]=0$. In our
case, one can realize $\langle J_{z}^{2}(t_0)\rangle=\xi_0^2\langle
J_{z}^{2}(0)\rangle$ with $\xi_0<1$, indicating the appearance of
the number-squeezed state \cite{Note}. Such a kind of squeezed state
has been observed in optical lattices \cite{Orzel}, optical trap
\cite{Chuu}, and atom chip \cite{Jo}. However, the observed
squeezing $\xi_0\simeq0.1$ for $N=4\times10^5$ \cite{Jo}, weaker
than our result $\xi_0\simeq1.467\times10^{-2}$ for
$\Omega_0=58.05\kappa$ and $N=2\times10^5$. Finally, within inset of
Fig.~\ref{fig4}, we show the time scale $t_{0}$ as a function of
$N$. Inserting $\Omega _{0}/\kappa =N^{1/3}$ into
Eq.~(\ref{t0_JLC}), we find that analytic expression of $t_{0}$
gives good agreement with the exact numerical simulations.

\section{Conclusion}

In summary, we have investigated optimal spin squeezing in a
two-component Bose-Einstein condensate with a Josephson coupling. We
show that: (i) the squeezing $\xi$ at time $t_0$ aligns along the
$z$ axis, which is equivalent with the number squeezing
\cite{Orzel,Schumm,Chuu,Jo,Note}, and is desirable for
high-precision atom interferometry \cite{Trimborn,phase}; (ii) there
exists a simple relation between the squeezing $\xi$ and the
single-particle coherence $g^{(1)}_{12}$ at $t_0$, Eq.(\ref{gq}) and
Eq.(\ref{gxi}), from which it is possible to measure the number
variance $\Delta J_z$ by readout the coherence $g^{(1)}_{12}$ in the
interference experiments; (iii) the strongest squeezing with the
power law $\xi_{0}\simeq 0.8578N^{-1/3}$ is achievable by applying
the optimal coupling $\Omega_0/\kappa\simeq N^{1/3}$. We also
discuss the maximal-squeezing time $t_0$ via the phase model and the
Husimi $Q$ function, and find that analytic result,
Eq.~(\ref{t0_JLC}), agrees with its numerical simulations.

\begin{acknowledgments}
We thank Professor C.~P.~Sun, Professor W.~M.~Liu, and Professor
S.~W.~Kim for helpful discussions. This work is supported by the
Research Grants Council of Hong Kong, Special Administrative Region
of China (Project No.~401406). GRJ acknowledge additional support by
the NSFC (Project No.~10804007), and Research Funds of Beijing
Jiaotong University (No.~2007RC030 and No.~2007XM049).
\end{acknowledgments}



\begin{thebibliography}{99}
\bibitem{Kitagawa} M. Kitagawa and M. Ueda, Phys. Rev. A \textbf{47}, 5138
(1993).

\bibitem{Wineland} D. J. Wineland \textit{et al.}, Phys. Rev. A \textbf{46},
R6797 (1992); \textit{ibid} \textbf{50}, 67 (1994).

\bibitem{Kuzmich} J. Hald \textit{et al.}, Phys. Rev. Lett. \textbf{83}, 1319 (1999);
A. Kuzmich \textit{et al.}, \textit{ibid} \textbf{85}, 1594 (2000);
J. M. Geremia \textit{et al.}, Science \textbf{304}, 270 (2004).

\bibitem{You} K. Helmerson and L. You, Phys. Rev. Lett. \textbf{87}, 170402
(2001); M. Zhang \textit{et al.}, Phys. Rev. A \textbf{68}, 043622
(2003); X. Wang and B. C. Sanders, Phys. Rev. A \textbf{68}, 012101
(2003); S. Yi and H. Pu, Phys. Rev. A \textbf{73}, 023602 (2006); I.
Tikhonenkov \textit{et al.}, Phys. Rev. A \textbf{77}, 063624
(2008).



\bibitem{Sorensen} A. S\o rensen \textit{et al.}, Nature (London) \textbf{409%
}, 63 (2001); U. V. Poulsen and K. Molmer, Phys. Rev. A \textbf{64}, 013616
(2001); Y. Li \textit{et al.}, Phys. Rev. Lett. \textbf{100}, 210401 (2008).

\bibitem{Takeuchi} M. Takeuchi \textit{et al.}, Phys. Rev. Lett. \textbf{94}%
, 023003 (2005).


\bibitem{Lewenstein} M. Lewenstein and L. You, Phys. Rev. Lett. \textbf{77},
3489 (1996); E. M. Wright \textit{et al.},\textit{ibid} \textbf{77},
2158 (1996); A. Imamo\={g}lu \textit{et al.}, \textit{ibid}
\textbf{78}, 2511 (1997); J. Javanainen and M. Wilkens,
\textit{ibid} \textbf{78}, 4675 (1997); Y. Castin and J. Dalibard,
Phys. Rev. A \textbf{55}, 4330 (1997); C. K. Law \textit{et al.},
\textit{ibid} \textbf{58}, 531 (1998).


\bibitem{Orzel} P. Bouyer and M.A. Kasevich, Phys. Rev. A {\bf 56}, R1083 (1997);
C. Orzel \textit{et al.}, Science \textbf{291}, 2386 (2001); M.
Greiner \textit{et al.}, Nature (London) \textbf{415}, 39 (2002); W.
Li \textit{et al.}, Phys. Rev. Lett. \textbf{98}, 040402 (2007); M.
Rodr\'{i}guez \textit{et al.}, Phys. Rev. A {\bf 75}, 011601(R)
(2007).

\bibitem{Schumm} T. Schumm \textit{et al.}, Nature Phys. \textbf{1}, 57
(2005); Y. Shin \textit{et al.}, Phys. Rev. A \textbf{72}, 021604(R)
(2005).

\bibitem{Chuu} C.-S. Chuu \textit{et al.}, Phys. Rev. Lett. \textbf{95},
260403 (2005).

\bibitem{Jo} G.-B. Jo \textit{et al.}, Phys. Rev. Lett. \textbf{98}, 030407
(2007).

\bibitem{Artur} A. Widera \textit{et al.}, Phys. Rev. Lett. \textbf{100},
140401 (2008).

\bibitem{Vardi} Y. Khodorkovsky, G. Kurizki, and A. Vardi, Phys. Rev. Lett.
\textbf{100}, 220403 (2008).

\bibitem{Bigelow} S. Raghavan, H. Pu, P. Meystre, and N. P. Bigelow, Opt.
Commu. \textbf{188}, 149 (2001).

\bibitem{Law} C.K. Law, H.T. Ng, and P.T. Leung, Phys. Rev. A \textbf{63},
055601 (2001).

\bibitem{Jin07} G. R. Jin and S. W. Kim, Phys. Rev. Lett. \textbf{99},
170405 (2007); \textit{ibid}, Phys. Rev. A \textbf{76}, 043621
(2007).


\bibitem{Note} Here, spin squeezing along the $z$ axis means the suppresion of $\Delta
J_z$, which is equivalent with the number squeezing \cite{Orzel} due
to $\Delta J_z=\Delta \hat{n}_{i}$, where $\Delta
\hat{n}_{i}=\sqrt{\langle \hat{n}_{i}^2\rangle-\langle
\hat{n}_{i}\rangle^2}$ with $i=1,2$ are the number variances, and
$\hat{n}_{i}=\hat{a}_i^{\dagger}\hat{a}_i$ the number operators.


\bibitem{Hall} D. S. Hall \textit{et al.}, Phys. Rev. Lett. \textbf{81},
1539 (1998); \textit{idib}, 1543 (1998).

\bibitem{Stenger} J. Stenger \textit{et al.}, Nature \textbf{396}, 345
(1998).

\bibitem{TMA} G. J. Milburn \textit{et al.}, Phys. Rev. A \textbf{55}, 4318
(1997); A. Smerzi, S. Fantoni, S. Giovanazzi, and S. R. Shenoy,
Phys. Rev. Lett. \textbf{79}, 4950 (1997);  P. Villain \textit{et
al.}, J. Mod. Opt. {\bf 44}, 1775 (1997); J. I. Cirac \textit{et
al.}, Phys. Rev. A \textbf{57}, 1208 (1998); M.J. Steel and M. J.
Collett, \textit{ibid} \textbf{57}, 2920 (1998); D. Gordon and C. M.
Savage, \textit{ibid} \textbf{59}, 4623 (1999); L. M. Kuang and Z.
W. Ouyang, \textit{ibid} {\bf 61}, 023604 (2000); L. M. Kuang and L.
Zhou, {\it ibid} {\bf 68}, 043606 (2003); W. D. Li et al., Phys.
Rev. A {\bf 64}, 015602 (2001); P. Zhang et al., J. Phys. B {\bf
35}, 4647 (2002).


\bibitem{CSS} J. M. Radcliffe, J. Phys. A \textbf{4}, 313 (1971); F. T.
Arecchi \textit{et al.}, Phys. Rev. A \textbf{6}, 2211 (1972).

\bibitem{Hines} A.P. Hines, R.H. McKenzie, G.J. Milburn, Phys. Rev. A {\bf 67}, 013609
(2003).

\bibitem{Agarwal} G.S. Agarwal and R.R. Puri, Phys. Rev. A \textbf{39},
2969 (1989); G. R. Jin and W. M. Liu, \textit{ibid} \textbf{70}, 013803
(2004); G. R. Jin \textit{et al.}, J. Opt. B: Quantum Semiclass. Opt.
\textbf{6}, 296 (2004).


\bibitem{Gerbier} F.Gerbier et al., Phys. Rev. Lett.{\bf 96}, 090401 (2006).

\bibitem{Pegg}D.T. Pegg and S.M. Barnett, Phys. Rev. A{\bf 39}, 1665 (1989).


\bibitem{phase model} D. Jaksch \textit{et al.},
Phys. Rev. Lett. \textbf{86}, 4733 (2001); C. Menotti \textit{et al.},
Phys. Rev. A \textbf{63}, 023601 (2001); A. Micheli \textit{et al.}, \textit{%
ibid}. \textbf{67}, 013607 (2003).

\bibitem{Leggett} A.J. Leggett, Rev. Mod. Phys. \textbf{73}, 307 (2001).

\bibitem{Trimborn} F. Trimborn \textit{et al.}, arXiv: quant-ph/0802.1142.

\bibitem{phase} C.Lee, Phys. Rev. Lett. {\bf 97} 150402 (2006); C.Lee
\textit{et al.}, Euro. Phys. Lett. {\bf 81} 60006 (2008); Y. P.
Huang and M. G. Moore, Phys. Rev. Lett. \textbf{100}, 250406 (2008).
\end{thebibliography}
\end{document}